\documentclass[final,1p,times]{elsarticle}
\usepackage{graphicx}
\usepackage{dcolumn}
\usepackage{bm}
\usepackage[english]{babel}
\usepackage{t1enc}
\usepackage{subfigure}
\usepackage{times}

\journal{Astroparticle Physics}

\begin{document}
\begin{frontmatter}
\title{Search for signatures of magnetically-induced alignment in the arrival directions measured by the Pierre Auger Observatory}
\author{ {\bf The Pierre Auger Collaboration} \\
P.~Abreu$^{74}$,
M.~Aglietta$^{57}$,
E.J.~Ahn$^{93}$,
I.F.M.~Albuquerque$^{19}$,
D.~Allard$^{33}$,
I.~Allekotte$^{1}$,
J.~Allen$^{96}$,
P.~Allison$^{98}$,
J.~Alvarez Castillo$^{67}$,
J.~Alvarez-Mu\~{n}iz$^{84}$
% ...
}
\address{$^{1}$ Centro At\'{o}mico Bariloche and Instituto Balseiro (CNEA-
UNCuyo-CONICET), San Carlos de Bariloche, Argentina \\
% ...
$^{19}$ Universidade de S\~{a}o Paulo, Instituto de F\'{\i}sica, S\~{a}o
Paulo, SP, Brazil \\
% ...
$^{33}$ Laboratoire AstroParticule et Cosmologie (APC),
Universit\'{e} Paris 7, CNRS-IN2P3, Paris, France \\
% ...
$^{57}$ Istituto di Fisica dello Spazio Interplanetario (INAF),
 Universit\`{a} di Torino and Sezione INFN, Torino, Italy \\
% ...
$^{67}$ Universidad Nacional Autonoma de Mexico, Mexico, D.F.,
Mexico \\
% ...
$^{74}$ LIP and Instituto Superior T\'{e}cnico, Technical
University of Lisbon, Portugal \\
% ...
$^{84}$ Universidad de Santiago de Compostela, Spain \\
% ...
$^{93}$ Fermilab, Batavia, IL, USA \\
% ...
$^{96}$ New York University, New York, NY, USA \\
% ...
$^{98}$ Ohio State University, Columbus, OH, USA \\
% ...
\par\noindent
($\dagger$) Deceased \\
(a) at Konan University, Kobe, Japan \\}
\begin{abstract}
We present the results of an analysis of data recorded at the Pierre Auger Observatory in which we search for groups of directionally-aligned events (or `multiplets') which exhibit a correlation between arrival direction and the inverse of the energy. These signatures are expected from sets of events coming from the same source after having been deflected by intervening coherent magnetic fields. The observation of several events from the same source would open the possibility to accurately reconstruct the position of the source and also measure the integral of the component of the magnetic field orthogonal to the trajectory of the cosmic rays. We describe the largest multiplets found and compute the probability that they appeared by chance from an isotropic distribution. We find no statistically significant evidence for the presence of multiplets arising from magnetic deflections in the present data.
\end{abstract}
\end{frontmatter}
\section{Introduction}
The origin of ultra-high energy cosmic rays is a long-standing open question, and the identification of their sources is one of the primary motivations for the research conducted at the Pierre Auger Observatory.
If the density of cosmic rays sources is not too large, it is expected that there could be indications of the presence of multiplets, i.e. sets of events with different energy that come from a single point-like source. Due to the magnetic fields that cosmic rays traverse on their paths from their sources to the Earth, they will be deflected and this deflection is proportional to the inverse of their energy if the deflections are small. Therefore, to identify sets of cosmic rays that come from a single source, a search for events that show a correlation between their arrival direction and the inverse of their energy has been performed using the data recorded at the Pierre Auger Observatory. The observation of cosmic ray multiplets could allow for the accurate location of the direction of the source and could also provide a new means to probe the galactic magnetic field, as it should be possible to infer the value of the integral of the component of the magnetic field orthogonal to the trajectory of the cosmic rays. Note that to observe a correlated multiplet the source should be steady, in the sense that its lifetime is larger than the difference in the time delays due to the propagation in the intervening magnetic fields for the energies considered. Moreover, magnetic fields should also be steady in the same sense so that cosmic rays traverse approximately the same fields.\\

This study relies on the acceleration at the source of a proton component (or intermediate mass nuclei being accelerated and photo-disintegrated during extragalactic propagation with the deflections due to extragalactic magnetic fields being small compared to those in the Galaxy). Due to the magnitude of the known magnetic fields involved, heavy nuclei at these energies would appear spread over a very large region of the sky, probing regions with different amplitudes and directions of the magnetic field, and hence losing their alignment and correlation with the inverse of energy.\\

The galactic magnetic field is poorly constrained by the available data, even though there has been considerable effort to improve this knowledge using different observational techniques, see e.g. \cite{han,beck,brown}. This field is usually described as the superposition of a large-scale regular component and a turbulent one. The regular component has a few $\mu$G strength and is coherent on scales of a few kpc with a structure related to the spiral arms of the galactic disk, and eventually also a more extended halo component (see e.g. \cite{pshirkov}). The deflection of cosmic rays with energy $E$ and charge $Z$ by the regular component of the magnetic field $\vec{B}$ after traversing a distance $L$ is given by
\begin{equation}
\delta \simeq 16^\circ \frac{20\ {\rm EeV}}{E/Z} \left | \int_{0}^{L}
       {\rm\frac{d \vec{l}}{3\ kpc} \times \frac{\vec{B}}{2\ \mu
           {\rm G}}}\right |,
\label{delta}
\end{equation}
where 1~${\rm EeV} \equiv 10^{18}$~eV.
This is the predominant deflection because, although the turbulent component has a root mean square amplitude of $B_{\rm rms} \simeq (1-2) B_{\rm reg}$, it has a much smaller coherence length (typically $L_c \simeq 50$-100 pc) \cite{rand,ohno}, leading to a smaller deflection, with a typical root mean square value
\begin{equation}
\delta_{\rm rms} \simeq 1.5^\circ \frac{20\ {\rm EeV}}{E/Z} \frac{{\rm B_{rms}}}{3\ \mu {\rm G}} \sqrt{\frac{L}{1\ {\rm kpc}}} \sqrt{\frac{L_c}{50\ {\rm pc}}}.
\end{equation}
After traveling a distance $L$ through the turbulent field, the trajectories of cosmic rays would be displaced a distance $\sim \delta_{\rm rms} L$ with respect to the one they would have had if only the regular field were present. If this displacement is smaller than the coherence length $L_c$, this means that all the particles with that energy have experienced nearly the same values of the turbulent field along their trajectories. Thus, the effect is that the arrival direction of cosmic rays will coherently wiggle with an amplitude $\delta_{\rm rms}(E)$ around the direction determined by the deflection due to the regular magnetic field as a function of the energy. Conversely, when $\delta_{\rm rms}(E) L > L_c $, particles of the same energy that have probed uncorrelated values of the turbulent field are able to reach the observer from the source and several images appear, scattered by $\delta_{\rm rms}(E)$ around the image that would be produced by the regular field alone. Which of the two regimes actually takes place depends on the energy considered and on the distance traveled in the turbulent field.
For instance, for $L\simeq 2$~kpc and energy about 20 EeV, the second situation
applies, while at much higher energies the first one holds.\\

Extragalactic magnetic fields could also deflect the trajectories of cosmic rays, but their strength is yet unknown and the relevance of their effect is a matter of debate, see e.g. \cite{sigl,dolag,das}.\\

\section{The Pierre Auger Observatory and the data set}

The Pierre Auger Observatory, located in Malargüe, Argentina ($35.2^{\circ}$S, $69.5^{\circ}$W) at 1400 m a.s.l. \cite{auger1}, was designed to measure ultra-high energy cosmic rays (energy $E>10^{18}$) with unprecedented statistics. It consists of a surface array of 1660 water-Cherenkov stations. The surface array is arranged in an equilateral triangular grid with 1500 m spacing, covering an area of approximately 3000 km$^2$ \cite{auger2}. The array is overlooked by 27 fluorescence telescopes located on hills at four sites on its periphery \cite{auger3}. The surface and air fluorescence detectors are designed to perform complementary measurements of air showers created by cosmic rays. The surface array is used to observe the lateral distribution of the air shower particles at ground level, while the fluorescence telescopes are used to record the longitudinal development of the shower as it moves through the atmosphere.\\

In this work we analyze events with zenith angles smaller than  $60^{\circ}$ recorded by the surface detector from 1st January 2004 to 31st December 2010. The events are required to have at least five active stations surrounding the station with the highest signal, and the reconstructed core must be inside an active
equilateral triangle of stations \cite{auger4}. The corresponding exposure is 25806 km$^2$ sr yr. The angular resolution, defined as the 68$^{\rm th}$ percentile of the distribution of opening angles between the true and reconstructed directions of simulated events, is better than $0.9^{\circ}$ for events that trigger at least six surface stations ($E > 10$ EeV) \cite{auger5}. The energy resolution is about $15\%$ and the absolute energy scale, given by the fluorescence calibration, has a systematic uncertainty of $22\%$ \cite{digiulio}.\\

\section{Method adopted for the multiplets search}

In the limit of large energy, and hence small deflections, it is a good approximation to consider the following simplified relation between the cosmic ray observed arrival directions, described by the unit vector $\vec{\theta}$, and the actual source direction $\vec{\theta_s}$
\begin{equation}
\vec{\theta}= \vec{\theta_s}+\frac{Ze}{E}\int_{0}^{L} {\rm d}\vec{l} \times \vec{B} \simeq \vec{\theta_s}+\frac{\vec{D}(\vec{\theta_s})}{E},
\end{equation}
where $Ze$ is the electric charge of the cosmic ray and $D\equiv|\vec{D}(\vec{\theta_s})|$ will be called the deflection power and will be given in units of 1$^{\circ}$ 100 EeV, which is $\approx$ 1.9 $e$ $\mu$G kpc.\\

In the case of proton sources, departures from the linear approximation are relevant for energies below 20 EeV for typical galactic magnetic field models \cite{multipletes}, as the deflections of the trajectories are large and the integral of the magnetic field component orthogonal to the path cannot be approximated as a constant for a fixed source direction. This fact motivates the restriction of the present analysis to events with energies above 20~EeV.\\

In order to identify sets of events coming from the same source, the main requirement will be that they appear aligned in the sky and have a high value of the correlation coefficient between the arrival direction and the inverse of the energy.\\

\begin{figure}[!htb]
\subfigure[\label{fig1a}]{\includegraphics[scale=0.44]{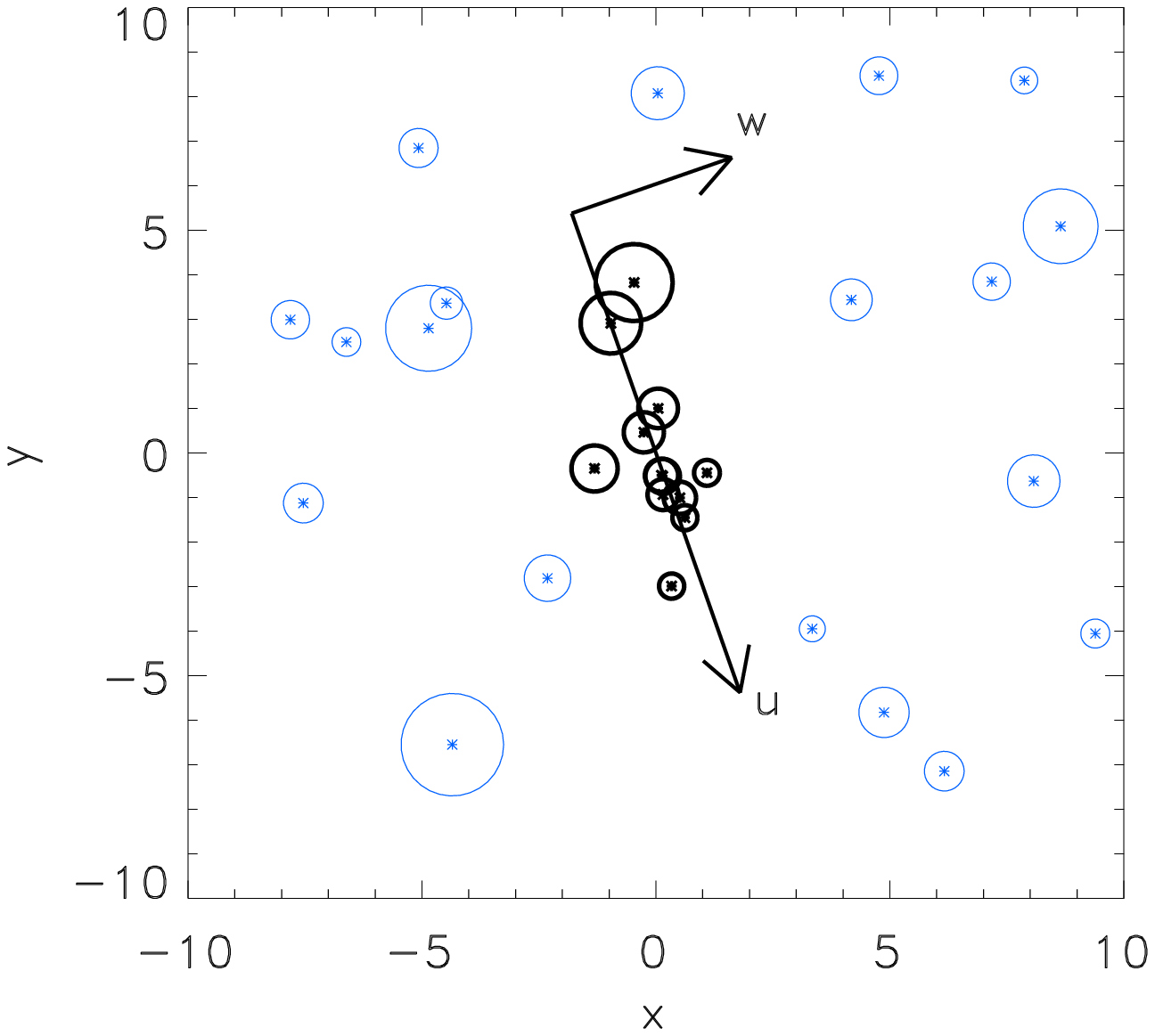}}
\subfigure[\label{fig1b}]{\includegraphics[scale=0.5]{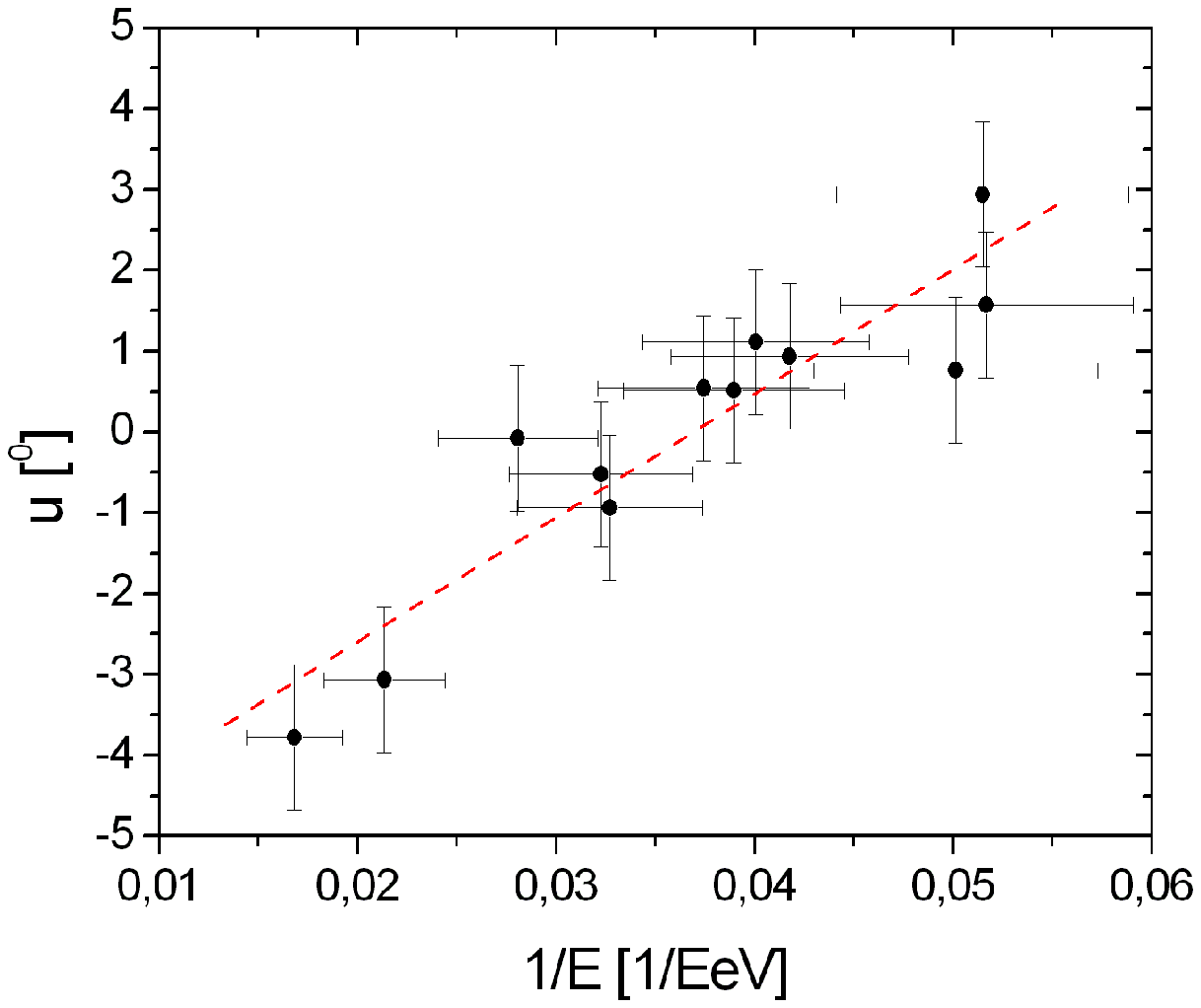}}
\caption[]{Selection of coordinates $u$ and $w$ for a set of events of a simulated source (black thick asterisks) superimposed on a background of isotropically-distributed events (blue asterisks) (a). The size of the circles is proportional to the energy of the events. Correlation between $u$ and $1/E$ for the same source events (b).} \label{fig1}
\end{figure}

To compute the correlation coefficient for a given subset of $N$ nearby event directions, we first identify the axis along which the correlation is maximal. For this we initially use an arbitrary coordinate system $(x,y)$ in the tangent plane to the celestial sphere (centered in the average direction to the events) and compute the covariance
\begin{equation}
{\rm Cov}(x,1/E)= \frac{1}{N} \sum _{i=1}^{N} (x_i- \left <x \right >
)(1/E_i- \left < 1/E \right >)
\end{equation}
and similarly for ${\rm Cov}(y,1/E)$. We then rotate the coordinates to a system ($u,w$) in which ${\rm Cov}(w,1/E)=0$, and hence ${\rm Cov}(u,1/E)$ is maximal. This corresponds to a rotation angle between the $u$ and $x$ axes given by
\begin{equation}
\alpha= \arctan \left ( \frac{{\rm Cov}(y,1/E)}{{\rm Cov}(x,1/E)} \right ).
\end{equation}
The correlation between $u$ and $1/E$ is measured through the correlation coefficient
\begin{equation}
C(u,1/E)=\frac{{\rm Cov}(u,1/E)}{\sqrt{{\rm Var}(u){\rm Var}(1/E)}},
\end{equation}
where the variances are given by ${\rm Var}(x)= \left < \left ( x-\left < x \right > \right ) ^2 \right >$. We demonstrate this procedure in Figure \ref{fig1}. In the left panel we show the selection of coordinates $u$ and $w$ for a set of events of a simulated source superimposed on a background of isotropically distributed events. In the right panel the correlation between $u$ and $1/E$ for the same source events is plotted. \\

\begin{figure}[!htb]
\subfigure[\label{fig2a}]{\includegraphics[scale=0.47]{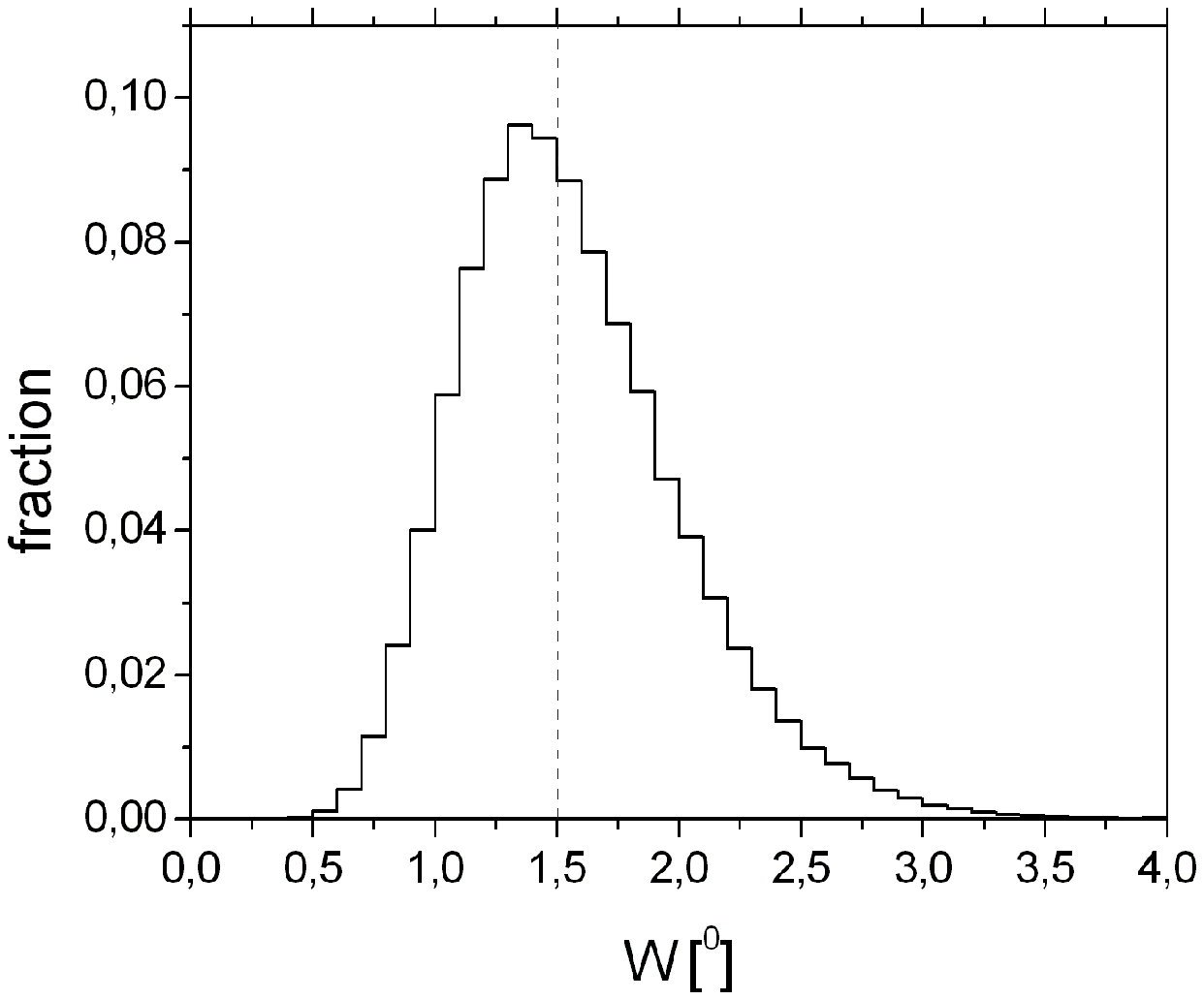}}
\subfigure[\label{fig2b}]{\includegraphics[scale=0.47]{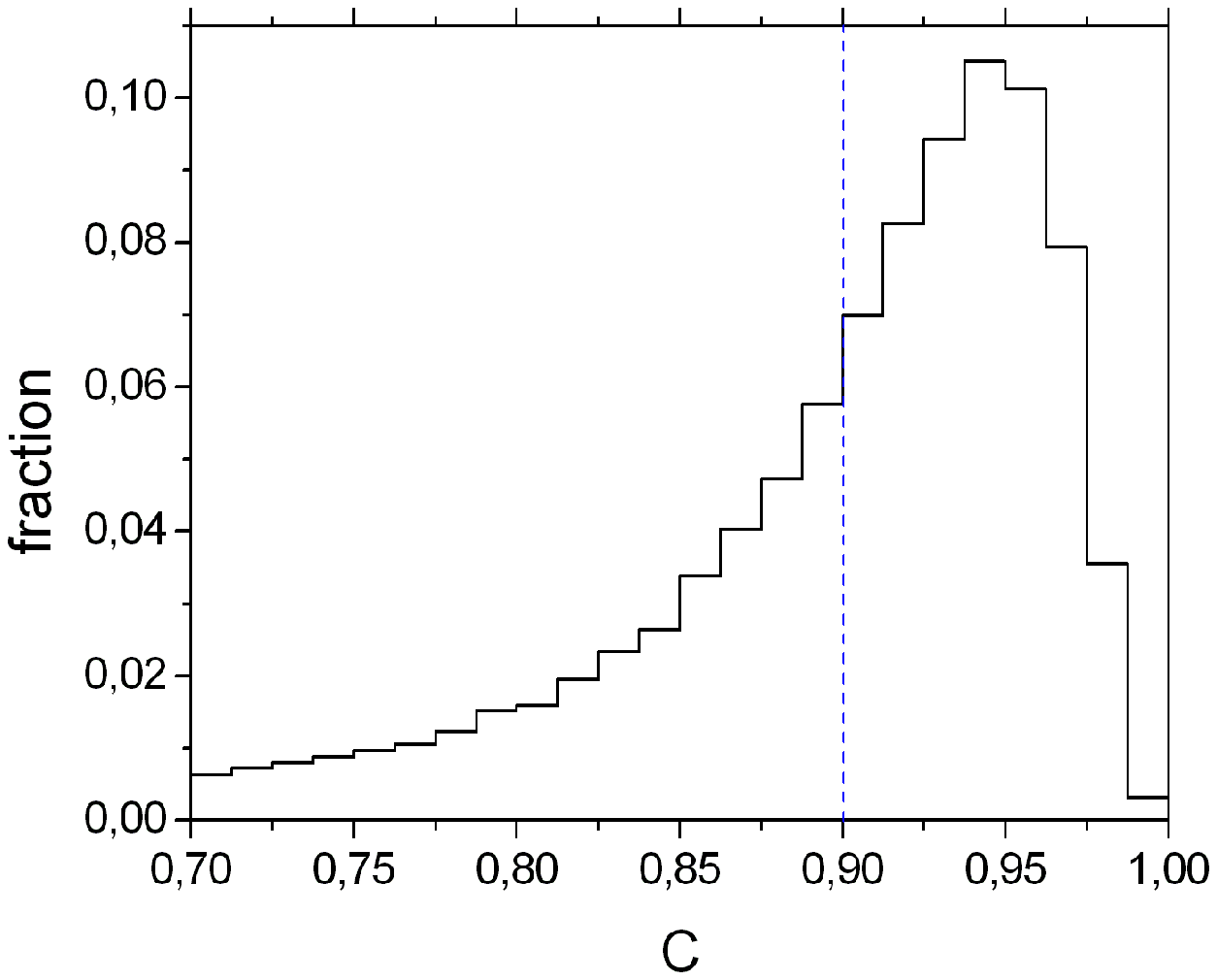}}
\caption[]{Distribution of the maximum angular distance $W$ (a) and the correlation coefficient $C(u,1/E)$  between the angular position $u$ and $1/E$ (b) for 14-plets from the 100 simulated sources. The vertical dashed lines indicate the cuts on $W$ and $C$ optimized for multiplicity and significance (see text).}
\end{figure}

\begin{figure}[!htb]
\subfigure[\label{fig3a}]{\includegraphics[scale=0.47]{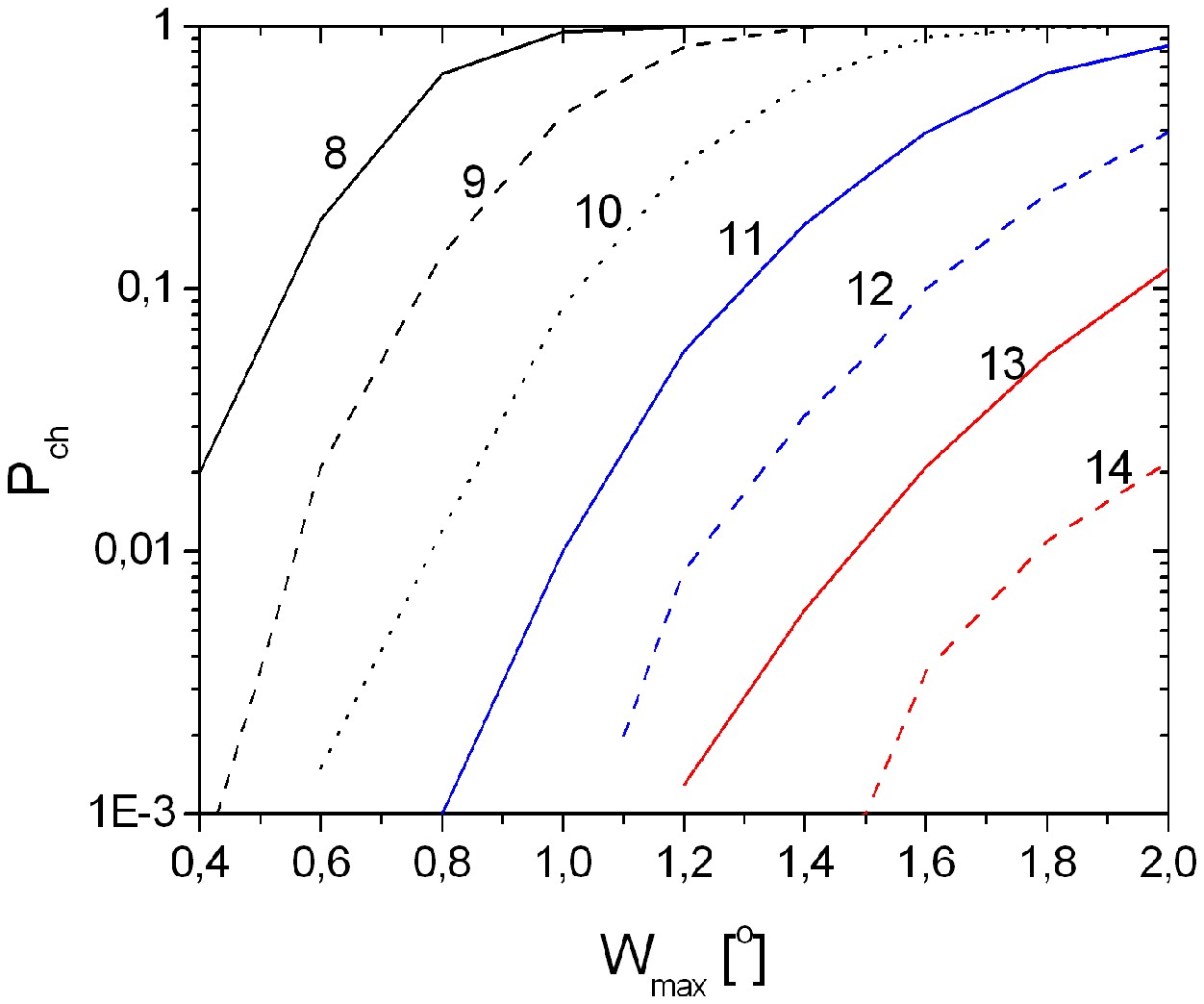}}
\subfigure[\label{fig3b}]{\includegraphics[scale=0.47]{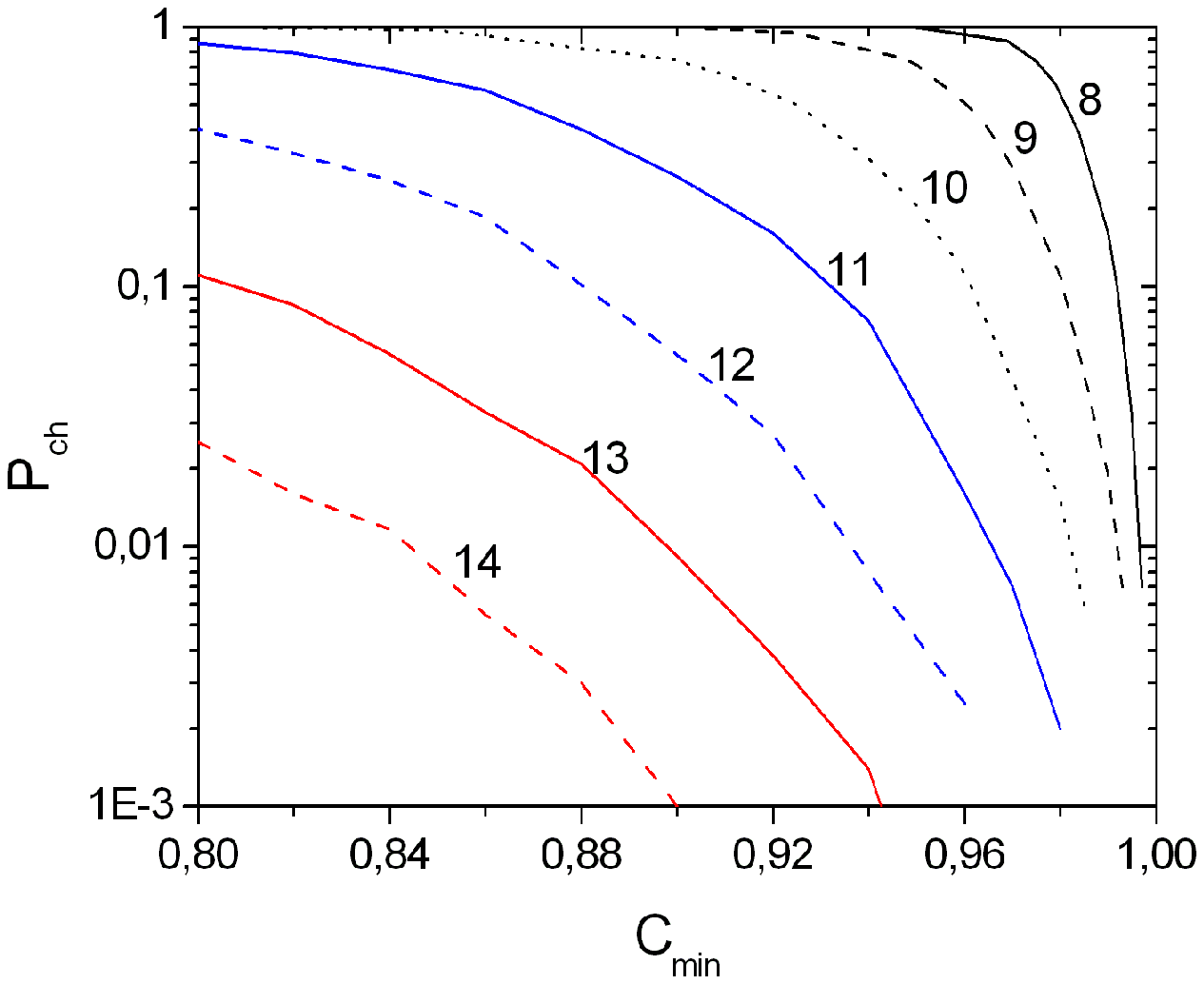}}
\caption[]{Chance probability $P_{\rm ch}$ for finding in isotropic simulations one large multiplet of a given multiplicity as a function of $W_{\rm max}$ (adopting $C_{\rm min}=0.9$) (a) and as a function of $C_{\rm min}$ (adopting $W_{\rm max}=1.5^\circ$) (b) (see text).}
\end{figure}

A given set of events will be identified as a correlated multiplet
when $C(u,1/E) > C_{\rm min}$ and, when the spread in the transverse
direction $w$ is small, $ W =\max(| w_i -\left < w \right > |) <
W_{\rm max}$ (corresponding to a total width of $\sim 2  W_{\rm max}$
in the perpendicular direction). The values for $C_{\rm min}$ and
$W_{\rm max}$ were chosen as a compromise between maximizing the
signal from a true source and minimizing the background arising from
chance alignments. In order to determine the optimal values of these
quantities, we  performed numerical simulations of sets of events from
randomly-located extragalactic sources. In these simulations, protons
were propagated through a bisymmetric magnetic field with even
symmetry (BSS-S) \cite{stanev,toes} (the local value of the field used
was $2\ \mu$G) and the effect of the turbulent magnetic field was
included by simply adding a random deflection with root mean square
amplitude $\delta_{\rm rms}=1.5^\circ(20\ {\rm EeV}/E)$. Although the
latter is a rough approximation, and a dependence on the arrival
directions should be expected, it is good enough for the purpose of
fixing $C_{\rm min}$ and $W_{\rm max}$. We considered one hundred
extragalactic sources located at random isotropic directions and
simulated sets of $N$ events coming from each source
($N=14,13,12$). The energy of the events followed an $E^{-2}$ spectrum
at the source and we added random gaussian uncertainties in the
angular directions and energies to account for the experimental
resolution. Magnetic lensing effects \cite{toes} were taken into
account in the simulation through the magnification or demagnification
of the energy spectrum of each source. As an example we show in Figure
\ref{fig2a} the resulting distribution of $W$ for multiplets of 14
events. The significance of a given multiplet can be quantified by
computing the fraction of  isotropically distributed simulations, with
the same total number of events as in the data and with the same
energy spectrum, in which a multiplet with the same or larger
multiplicity and passing the same cuts appears by chance. At high
energies the UHECR angular distribution may not be isotropic,
reflecting structure in the distribution of sources within the GZK
horizon. However, our data set is dominated by lower energy events for
which isotropy is an excellent approximation. We show in Figure
\ref{fig3a} the chance probability for multiplets of different
multiplicity as a function of $W_{\rm max}$. We note that when
reducing $W_{\rm max}$, some of the events of the multiplets will be
missed and their multiplicity will be reduced. However, the
significance of a smaller multiplet passing a tighter bound on $W_{\rm
  max}$ can be larger than the significance of the complete multiplet
with a looser $W_{\rm max}$ cut. It turns out that the largest mean
significance for the simulated sources (i.e. the average of the
significances of the resulting multiplets after imposing the cuts)
appears when a cut $W_{\rm max} \simeq 1.5^{\circ}$ is applied.
The angular scale
of $1.5^\circ$ provides in fact a reasonable cut which accounts for
the angular resolution and the mean value of the turbulent field
deflections. We note that in the case of 14-plets, in $50\%$ of
the simulations all the events pass this cut and the multiplet will be
reconstructed as a 14-plet, while in $38\%$ of the cases one event is
lost and in $11\%$ of the cases two events are lost. \\

A similar analysis can be performed to fix the cut on the correlation
coefficient $C_{\rm min}$. The distribution of $C(u,1/E)$ for the
simulated 14-plets is shown in Figure \ref{fig2b} and the chance
probability for multiplets of different multiplicity as a function of
$C_{\rm min}$ is illustrated in Figure \ref{fig3b}. The largest mean
significance is attained now for values of $C_{\rm min}$ in the range
from 0.85 to 0.9, depending on the multiplicity considered. We will
then fix in the following
$W_{\rm max}= 1.5^\circ$ and $C_{\rm min}=0.9$. Considering
simulations with 14 events  and for a cut
$C_{\rm min}=0.9$, we find that in $57\%$ of the cases all events
pass the cuts, in $12\%$ of the simulations one event is lost and in
$11\%$ of them two events are lost.  We note that the
choice of the optimal cuts depends slightly on the galactic magnetic
field model considered in the simulations and on the modeling of the
turbulent field deflections.\\

We will hence search for correlated multiplets of events with energies above
20~EeV (so that the linear correlation of the deflection with $1/E$ is
still expected to be valid for proton sources) which
extend up to $20^{\circ}$ in the sky (see eq.~(\ref{delta})). We also
require that the multiplet contains at least one event with energy
above 45~EeV. We note that the energy of the
most energetic event of a set of 10 events with $E>20$~EeV from a
source with spectral index $s=2.5$ is larger than 45~EeV with a
probability of $97\%$ (for a spectral index $s=3$ this probability is
$\sim 90\%$ and for $s=2$ it is $\sim 99.7\%$ ). Hence, requiring one
high energy event above 45~EeV is not restrictive, and it simplifies
the strategy to start the search for multiplets, which proceeds by
looking at all possible sets of events contained in windows of
$20^\circ$ around those high energy events. Since we are ultimately
interested in multiplicities larger than 8 (see Fig. 3 in which it is
apparent that for the present statistics above 20~EeV correlated sets of smaller multiplicity are very likely to appear by chance in isotropic
simulations), it is possible to make this search more efficient by first
identifying the high energy end of the candidate multiplets.
We hence consider for every event above 45~EeV the quadruplets that
it forms with the events within a circle of $15^\circ$ having energies above
25~EeV and with a correlation coefficient $C(u,1/E) \ge 0.8$. The precise values of these cuts are not crucial as long as they allow one to safely include the larger multiplets of interest. For each
of these candidates we then extend the search including all the events above
20~EeV with an angular distance to the highest energy one smaller than
$20^\circ$ and at a distance smaller than $3 W_{max}$ from the quadruplet axis.
This allows us to find the correlated multiplets satisfying the cuts in $W_{max}$
and $C_{min}$ in a very efficient way, as it is desirable to be able to perform
a large number of simulations.\\

The multiplets search procedure has been designed for sources
having a light composition. For sources having
instead a heavy composition above 20~EeV, multiplets will be much more
difficult to identify since they would typically spread through a
larger region in the sky and also the linearity of their directional distribution will be
lost.\\

Once a correlated multiplet is identified, from the linear fit to the relation
\begin{equation}
u= u_s + \frac{D}{E},
\end{equation}
the position of the source ($u_s,0$) (in the $u$-$w$ coordinate
system) and the deflection power $D$ can be obtained.\\

A true correlated multiplet arising from magnetic field deflections of
events from a single source can also include by chance some events from
the background that appear aligned and correlated in energy with the
events from the source. We have estimated the fraction of events that
is expected to be due to chance background alignments by simulating an
isotropic background distribution of events with the energy of the
observed events above 20 EeV and superimposing multiplets of 12 events
from simulated sources. We found that $29\%$ of the reconstructed
multiplets do not pick additional background events, while $46\%$ just
pick one additional background event and $25\%$ pick two or
more. Thus, the fraction of events added from the background is
typically very small.\\

\section{Results}

\begin{figure}[!h]
\begin{center}
\includegraphics[scale = 1]{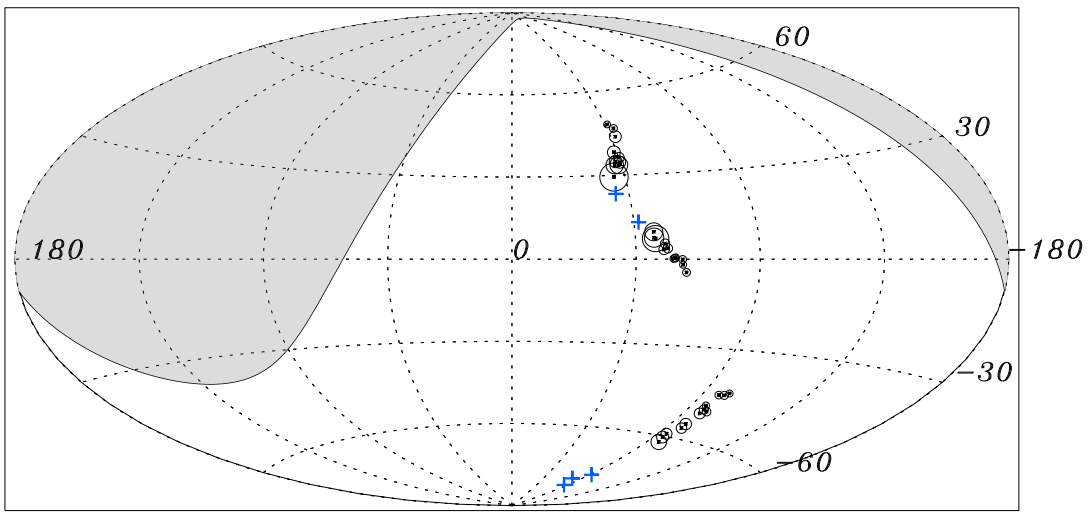}
\caption[]{Observed multiplets with 10 or more events in galactic coordinates. The size of the circles is
proportional to the energy of the event. Plus signs indicate the positions of the potential sources for each multiplet. One decuplet is in fact three dependent decuplets that are formed by the combination of twelve events and the three corresponding reconstructions of the potential sources are shown. The solid line represents the border of the field of view of the Southern Observatory for zenith angles smaller than $60^\circ$ and the grey shaded area is the region outside the field of view.}\label{fig4}
\end{center}
\end{figure}

We applied the method discussed in Section 3 to 1509 events above 20
EeV recorded at the Pierre Auger Observatory from 1st January 2004 to
31st December 2010. We implemented a search for all possible
multiplets which extend up to 20$^\circ$ in the sky and contain at
least one event with energy above 45 EeV, and that have a half-width
smaller than $W_{\rm max}=1.5^\circ$ and a correlation coefficient
larger than $C_{\rm min}=0.9$. The largest multiplet found in this
data set is one 12-plet and there are also two independent
decuplets. They are displayed in Figure \ref{fig4}. Their deflection
power, position of the potential source location and correlation
coefficient are listed in Table \ref{tabla}. Decuplet II in Table
\ref{tabla} consists of three dependent sets of ten events (a-c) that
are formed by the combination of a set of twelve events. These three
decuplets are not independent of each other since they have most
events in common. The uncertainties in the reconstruction of the
position of the potential sources have been calculated propagating the
uncertainties in energy and arrival direction to an uncertainty in the
rotation angle (Eq.~5) and in the linear fit performed to the
deflection vs. $1/E$ (Eq.~7).\\

The probability that the observed number (or more) of correlated multiplets appears by chance can be computed by applying a similar analysis to simulations of randomly distributed events weighted by the geometric exposure of the experiment \cite{sommers} and with the energies of the observed events. The fraction of simulations with at least one multiplet with 12 or more events is $6\%$, and the fraction having at least three multiplets with 10 or more events is $20\%$. Therefore, there is no statistically significant evidence for the presence of multiplets from actual sources in the data. We note that with the present statistics, an individual multiplet passing the required selection cuts should have at least 14 correlated events in order that its chance probability be $10^{-3}$.\\

Measurements by the Pierre Auger Observatory \cite{xmax} of the depth
of shower maximum and its fluctuations indicate a trend towards heavy
nuclei with increasing energy.  This interpretation of the shower
depths is not certain, however. It relies on shower simulations that
use hadronic interaction models to extrapolate particle interaction
properties two orders of magnitude in centre-of-mass energy beyond the
regime where they have been tested experimentally. Magnetic alignment
and correlation with the inverse of the energy as searched here are
not expected for heavy nuclei. Assuming there are sources which
accelerate an appreciable proton component, the non-observation of
significant multiplets could be the consequence of having a large
density of sources. Given the present statistics, the maximum source
density which would allow to observe a multiplet containing 12 events
above 20 EeV from the nearest source to the Earth can be roughly
estimated by considering that this source should produce a fraction
$12/1509 \approx 1/125$ of the total flux observed in the field of
view of the Auger Observatory in this energy range. Assuming that the
sources have equal intrinsic luminosity and are uniformly distributed
and that cosmic rays in this energy range can arrive from distances up
to about 1 Gpc, the above mentioned constraints imply that the nearest
source should be within $\sim 10$ Mpc. Thus, the mean local density of
sources should not be larger than a few $10^{-4}$ Mpc$^{-3}$. The fact
that we have not seen a larger multiplet is an indication that the
density of sources is probably larger.
This very rough estimation is subject to large
fluctuations but it is indicative that densities within the current
lower limits may lead to the kind of signals searched for here.
We note, however, that this
bound would be relaxed if contributions of heavy cosmic ray primaries
become significant, or if very strong turbulent magnetic fields were present.\\

\begin{table}[!hbtp]
\begin{tabular}{l|ccccc}
\hline \hline
\\
${\rm multiplet}$~~    & ~~ $D [^{\circ} 100\ {\rm EeV}]$~~ & ~~$(l,b)_S [^{\circ}]$ & ~~ $\Delta u_S [^{\circ}]$ & ~~ $\Delta w_S [^{\circ}]$ & ~~ $C$~~\\
\\
\hline
\\
${\rm 12-plet}$~~         & ~~$4.3 \pm 0.7$~~     & ~~$(-46.7,13.2)$~~   & ~~$2.4$~~ & ~~$0.9$~~ & ~~$0.903$~~\\
${\rm 10-plet\ I}$~~      & ~~$5.1 \pm 0.9$~~     & ~~$(-39.9,23.4)$~~   & ~~$2.7$~~ & ~~$0.9$~~ & ~~$0.901$~~\\
${\rm 10-plet\ IIa}$~~    & ~~$8.2 \pm 1.3$~~     & ~~$(-85.6,-80.4)$~~  & ~~$4.3$~~ & ~~$1.9$~~ & ~~$0.920$~~\\
${\rm 10-plet\ IIb}$~~    & ~~$7.6 \pm 1.2$~~     & ~~$(-79.6,-77.9)$~~  & ~~$4.0$~~ & ~~$1.6$~~ & ~~$0.919$~~\\
${\rm 10-plet\ IIc}$~~    & ~~$6.5 \pm 1.1$~~     & ~~$(-91.5,-75.7)$~~  & ~~$3.9$~~ & ~~$1.6$~~ & ~~$0.908$~~\\
\\
\hline \hline
\end{tabular}
\caption{Deflection power, $D$; reconstructed position of the potential source in galactic coordinates, $(l,b)_S$; uncertainty in the reconstructed position of the potential source along the direction of deflection, $\Delta u_S$, and orthogonal to it, $\Delta w_S$; and linear correlation coefficient, $C$, for the largest correlated multiplets found. The data correspond to events with energy above 20 EeV from 1st January 2004 to 31st December 2010. \label{tabla}}
\end{table}

\section{Conclusions}

A search for ultra-high energy cosmic ray multiplets was performed in the data gathered between 1st January 2004 and 31st December 2010 by the Pierre Auger Observatory with energy above 20 EeV. The largest multiplet found was one 12-plet. The probability that it appears by chance from an isotropic distribution of events is $6\%$. Thus, there is no significant evidence for the existence of correlated multiplets in the present data set. Future data will be analyzed to check if some of the observed multiplets grow significantly or if some new large multiplets appear.\\

\section{Acknowledgments}
The successful installation, commissioning and operation of the Pierre Auger Observatory
would not have been possible without the strong commitment and effort
from the technical and administrative staff in Malarg\"ue.

We are very grateful to the following agencies and organizations for financial support:
Co\-misi\'on Nacional de Energ\'ia At\'omica,
Fundaci\'on Antorchas,
Gobierno De La Provincia de Mendoza,
Municipalidad de Malarg\"ue,
NDM Holdings and Valle Las Le\~nas, in gratitude for their continuing
cooperation over land access, Argentina;
the Australian Research Council;
Conselho Nacional de Desenvolvimento Cient\'ifico e Tecnol\'ogico (CNPq),
Financiadora de Estudos e Projetos (FINEP),
Funda\c{c}\~ao de Amparo \`a Pesquisa do Estado de Rio de Janeiro (FAPERJ),
Funda\c{c}\~ao de Amparo \`a Pesquisa do Estado de S\~ao Paulo (FAPESP),
Minist\'erio de Ci\^{e}ncia e Tecnologia (MCT), Brazil;
AVCR AV0Z10100502 and AV0Z10100522,
GAAV KJB100100904,
MSMT-CR LA08016, LC527, 1M06002, and MSM0021620859, Czech Republic;
Centre de Calcul IN2P3/CNRS,
Centre National de la Recherche Scientifique (CNRS),
Conseil R\'egional Ile-de-France,
D\'epartement  Physique Nucl\'eaire et Corpusculaire (PNC-IN2P3/CNRS),
D\'epartement Sciences de l'Univers (SDU-INSU/CNRS), France;
Bundesministerium f\"ur Bildung und Forschung (BMBF),
Deutsche Forschungsgemeinschaft (DFG),
Finanzministerium Baden-W\"urt\-tem\-berg,
Helmholtz-Gemeinschaft Deutscher Forschungszentren (HGF),
Ministerium f\"ur Wissenschaft und Forschung, Nordrhein-Westfalen,
Ministerium f\"ur Wissenschaft, Forschung und Kunst, Baden-W\"urt\-tem\-berg, Germany;
Istituto Nazionale di Fisica Nucleare (INFN),
Ministero dell'Istruzione, dell'Universit\`a e della Ricerca (MIUR), Italy;
Consejo Nacional de Ciencia y Tecnolog\'ia (CONACYT), Mexico;
Ministerie van Onderwijs, Cultuur en Wetenschap,
Nederlandse Organisatie voor Wetenschappelijk Onderzoek (NWO),
Stichting voor Fundamenteel Onderzoek der Materie (FOM), Netherlands;
Ministry of Science and Higher Education,
Grant Nos. 1 P03 D 014 30, N202 090 31/0623, and PAP/218/2006, Poland;
Funda\c{c}\~ao para a Ci\^{e}ncia e a Tecnologia, Portugal;
Ministry for Higher Education, Science, and Technology,
Slovenian Research Agency, Slovenia;
Comunidad de Madrid,
Consejer\'ia de Educaci\'on de la Comunidad de Castilla La Mancha,
FEDER funds,
Ministerio de Ciencia e Innovaci\'on and Consolider-Ingenio 2010 (CPAN),
Xunta de Galicia, Spain;
Science and Technology Facilities Council, United Kingdom;
Department of Energy, Contract Nos. DE-AC02-07CH11359, DE-FR02-04ER41300,
National Science Foundation, Grant No. 0450696,
The Grainger Foundation USA;
ALFA-EC / HELEN,
European Union 6th Framework Program,
Grant No. MEIF-CT-2005-025057,
European Union 7th Framework Program, Grant No. PIEF-GA-2008-220240,
and UNESCO.

\end{document}